\documentclass[journal=jacsat,manuscript=article,layout=twocolumn]{achemso}
\usepackage[version=3]{mhchem} 
\usepackage{color}
\usepackage{setspace}\usepackage{threeparttable}
\usepackage{upgreek}

\author{M. Lessel}

\author{O. B\"aumchen}
\altaffiliation{Present address: Department of Physics and Astronomy, McMaster University, Hamilton, ON, Canada L8S 4M1}

\author{M. Klos}
\author{H. H\"ahl}
\altaffiliation{Present address: University of Zurich, Institute of Physical Chemistry, Winterthurerstrasse 190, 8057 Zurich, Switzerland}
\author{R. Fetzer}
\altaffiliation{Present address: Karlsruhe Institute of Technology, Institute for Pulsed Power and Microwave
Technology, 76344 Eggenstein-Leopoldshafen, Germany}

\author{R. Seemann}

\author{K. Jacobs}
\email{k.jacobs@physik.uni-saarland.de}
\affiliation{Department of Experimental Physics, Saarland University, D-66041 Saarbr\"ucken, Germany}

\title{Self-assembled silane monolayers: A step-by-step high speed recipe for high-quality, low energy surfaces}

\begin{document}

\begin{abstract}
Silanization bases on the adsorption, self-assembly and covalent binding of silane molecules onto surfaces, resulting in a densely packed self-assembled monolayer (SAM). Following standard recipes, however, the quality of the monolayer is often variable and therefore unsatisfactory. The process of self-assembly is highly affected by the chemicals involved in the wet-chemical process, by the ambient parameters during preparation such as humidity or temperature and by possibly present contaminants (or impurities). Here, we present a reliable, efficient, and wet-chemical recipe for the preparation of ultra-smooth, highly ordered alkyl-terminated silane SAMs on \ce{Si} wafers.
\end{abstract}

\section{Introduction}
SAMs prepared on smooth surfaces like \ce{Si} wafers exhibit extraordinary properties such as chemical homogeneity, ultra low surface roughness and controlled wettability~\cite{Ulman1996,Schreiber2000,Ravoo2005}. The latter can be varied in wide ranges, depending on the end group of the silane. Here, we focus on alkyl-terminated silanes resulting in surfaces of very low surface energy. Silane layers in particular are mechanically robust~\cite{Ding2006} and thermally stable up to at least 250\,$^{\circ}$C~\cite{Srinivasan1998, Helmy2002}. These properties render silane-coated substrates ideal model surfaces to study a wide range of physical, chemical and biological phenomena such as adhesion~\cite{Prime1991,Prime1993,Faucheux2004}, adsorption~\cite{Cao2006,Tang2007,Haehl2012,Loskill2012}, friction~\cite{Tsukruk1998,Flater2007,Chandross2008,Booth2011} or nanofluidics of thin liquid films~\cite{Pit2000,Cottin2002,Fetzer2005,Fetzer2006,Zhang2003,Zhu2009}. Alongside fundamental research, also technical applications benefit from their unique properties: SAMs act \textit{e.\,g.}\ as lubrication layers in micro-electro-mechanical-systems (MEMS)~\cite{Srinivasan1998} and as coatings in microfluidic devices~\cite{Tretheway2002,Tretheway2004}. 

\begin{figure}
  \includegraphics[width=0.4\textwidth]{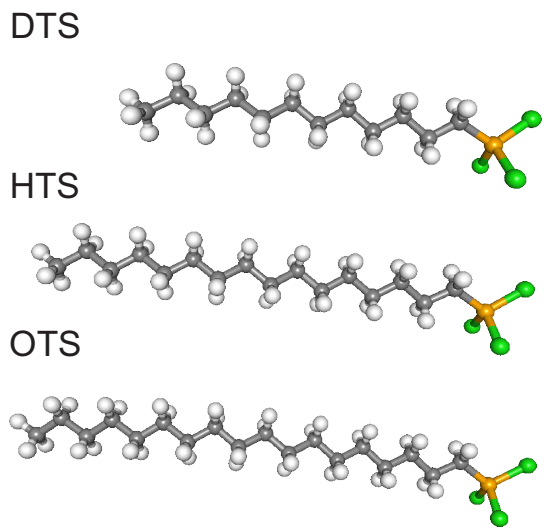}
  \caption{Ball and stick representation of alkylsilane molecules (white: hydrogen, gray: carbon, orange: silicon, green: chlorine). (Models illustrated using BALLView~\cite{Hildebrandt2010}.)}
  \label{Kalottenmodelle}
\end{figure}

The recipe presented here focuses on the widely-used \ce{CH3}-terminated alkylsilanes with three different chain lengths determined by the number of carbon atoms (12, 16 and 18) in the backbone: dodecyl-trichlorosilane (DTS, \ce{C12H25Cl3Si}), hexadecyl-trichlorosilane (HTS, \ce{C16H33Cl3Si}) and octadecyl-trichlorosilane (OTS, \ce{C18H37Cl3Si}), illustrated in Figure~\ref{Kalottenmodelle}.

The formation of self-assembled monolayers was first reported by Zisman~\cite{Bigelow1946,Bigelow1947} in the late 1940's. From the 1980's onwards, Sagiv and others~\cite{Sagiv1980,Wasserman1989,Schreiber2000,Banga1995} studied the underlying formation process and characterized surface properties like surface energy, layer thickness and roughness on alkylsilane SAMs. Meanwhile, a variety of experimental techniques have been applied to study SAMs: atomic force microscopy~\cite{Bierbaum1995}, ellipsometry~\cite{Sagiv1980,Wasserman1989}, contact angle measurements~\cite{Peters2002,Brzoska1994}, low frequency infrared spectroscopy~\cite{Tripp1995}, near edge X-ray absorption spectroscopy~\cite{Peters2002} or X-ray reflectometry~\cite{Mezger2008}. Most of these studies followed a wet-chemical preparation procedure, going back to the original work of Zisman~\cite{Bigelow1946,Bigelow1947}. It turns out that the quality of the silane SAM does not only depend on the recipe itself and on the quality of the used chemicals, but also on humidity, temperature and contaminants present during preparation~\cite{Brzoska1994,Peters2002,Wang2003,Wang2005}. In the following, we present a quick and easy-to-follow recipe to prepare silanized \ce{Si} wafers of reproducible quality, \textit{i.\,e.}\ of low roughness, high ordering, high water contact angle and low contact angle hysteresis.

\section{Preparation}
Alkylsilanes are offered from several suppliers in very high purity. For our study, DTS (> 99\,\%), OTS (> 90\,\%) and  HTS (95\,\%) were used (\textit{c.\,f.}\ table~\ref{Chemicals}). Flakes or filaments in the solution indicate aggregations of silane molecules and are typically caused by contact of the solution with water or humid air, inducing cross-linking of the silanes~\cite{Brzoska1994}. If this is the case, the solution should not be used. Silane solutions are packed under inert gas atmosphere from to supplier to prevent such cross-linking. We therefore recommend to use a new package of silane for each silanization. A complete list of neccessary chemicals is given in Table~\ref{Chemicals}.

\begin{table*}[h]
\small
  \caption{Chemicals necessary for silanization.}
  \label{Chemicals}
  \begin{tabular}{llll}
    \hline
    Chemicals&purity /or special product &CAS-no.&supplier \\
    \hline
  \underline{\textbf{cleaning:}}		&													&					&						\\
hydrogen peroxide 				& 30\,\%, non-stabilized\bibnote{It is essential to use non-stabilized \ce{H2O2}.}&7722-84-1 & Merck\\
sulphuric acid						&96\,\%, Selectipur\textsuperscript{\textregistered} &	7664-93-9 & BASF\\    
   	\underline{\textbf{solvents:}}       &                       &        &  \\ 
bicyclohexyl  						& 99\,\%								& 92-51-3 &Acros Organics\\
carbon tetrachloride			&  	&56-23-5 &local supplier used\\
chloroform								&> 99.8\,\%, LiChrosolv\textsuperscript{\textregistered} &	67-66-3 & Sigma-Aldrich\\

	\underline{\textbf{silanes:}}				&												&					&\\
octadecyl-trichlorosilane	& > 90\,\%& 112-04-9 &Sigma-Aldrich\\
hexadecyl-trichlorosilane	& 95\,\%&5894-60-0 &ABCR\\
dodecyl-trichlorosilane		&puriss. > 99\,\% &4484-72-4 &Fluka\\
   \hline
  \end{tabular}
\end{table*}

The coating process consists of four steps: (i) cleaning of all parts involved in the preparation, such as beakers and tweezers, (ii) cleaning of the \ce{Si} wafers, (iii) preparation of the solution for the silanization and (iv) the coating procedure itself~\bibnote{For handling the chemicals, local safety rules apply and the use of lab coats, goggles and suitable gloves is mandatory.}

\subsection{Cleaning Procedure}
A clean wafer surface is a precondition to achieve a covalent binding of the silane to the natural, amorphous \ce{SiO2} layer of the \ce{Si} wafer~\cite{Faidt2012}. Typically, organic residues left over from the polishing process of the silicon wafer have to be removed. Here, peroxymonosulfuric acid ('piranha solution')~\bibnote{A compact guide for handling 'piranha solution' can be found \textit{e.g.}\ at http://www.nanofab.ubc.ca/process-piranha-etch.}, consisting of 50\,\% \ce{H2SO4} and 50\,\% \ce{H2O2} has proven to be suitable as it does not roughen the wafer surface. For details of the chemicals see Table~\ref{Chemicals}.

All steps (i) - (iv) were performed in a dust-free environment, at least a class 100 clean room or equivalent environments (\textit{e.\,g.}\ a laminar flow work bench). All beakers~\bibnote{For the silanization of a 4\,inch wafer, three PTFE beakers of similar diameter~\bibnote{It is convenient to choose the diameter of the beaker only slightly larger than the diameter of the wafer as this allows a complete coverage of the wafer with a minimum amount of piranha solution and silane solution required.}, one 2\,l Erlenmeyer flask for the hot DI-water, one graduated cylinder (50\,ml) and pipettes for the chemicals of the silanization solution.} and tweezers to be used during the silanization process should withstand the cleaning procedure. Containers and tweezers made of Teflon\textsuperscript{\textregistered} or other PTFE polymers have proven to be suitable.  

For step (i), all beakers are filled with piranha solution. Then, the tweezers are given into the acid. To ensure complete cleaning, the beakers (together with the tweezers) are tilted and then slowly rotated by hand (5-10\,min), so that the piranha solution can reach the bezel of the beaker. Subsequently, beakers and tweezers are rinsed three times in hot deionized (DI-) water (18.2\,M$\Omega$cm at 25\,$^{\circ}$C and total-organic-carbon less than 6\,ppb). This routine is repeated twice, each with fresh piranha solution and fresh DI-water. To prevent new contamination, the tweezers are kept in the cleaned containers.

For step (ii), in the first PTFE beaker, one \ce{Si} wafer (with native \ce{SiO2}~\bibnote{Crystal orientation and type of doping of the \ce{Si}-wafers do not impact on the silanization}, \textit{e.g.}\ in our case purchased from Si-Mat Silicon Materials, Kaufering, Germany) is immersed (polished side up) into freshly prepared piranha solution straight from the dust-tight package of the wafer manufacturer. The formation of bubbles on the wafer surface is an indication of the acid's reactivity. A periodical, careful waving of the beaker removes bubbles from the wafer surface and stirs the liquid to ensure uniform oxidation of any organic residues on the surface. After 30\,min, the wafer is removed from the piranha solution and placed into the second PTFE beaker filled with hot DI-water. Promptly, the water is replaced by new hot DI-water. The wafer remains for about 10-20\,min in the new water. Then, the liquid is replaced every 20\,min, at least 3 times. This procedure is necessary to completely remove reaction products left over on the wafer surface. After the piranha treatment, the \ce{Si} wafer exhibits a very high surface energy and is thus susceptible to contamination. Therefore, the wafer should be kept in the DI-water until needed. While the first wafer is immersed in the DI-water a second wafer can cleaned with piranha according to step (ii).

\subsection{Preparation of the Silane Solution}
For step (iii), the third PTFE beaker, cleaned in step (ii), is thoroughly dried with an inert gas \textit{e.\,g.}\ \ce{N2} (5.0 grade)~\bibnote{Including an oil free blow gun}. The following quantities suffice for immersing one 4\,inch wafer at a time into a suitable beaker, requiring about 50\,ml of silane solution.
 
The basic solvent is bicyclohexyl. In 50\,ml of bicylohexyl, 30 drops (approx.\ 1.5\,$\upmu$ml) of carbon tetrachloride \ce{CCl4} and 15 drops (approx.\ 0.75\,$\upmu$l) of the silane are added. As \ce{CCl4} prevents the silane from forming micelles~\cite{Brzoska1994}, the solution is lucent. The appearance of white flakes or filaments indicates an uncontrolled cross-linking of silane molecules as described above. In this case, the solution should not be used. We achieved the presented results at relative humidities below 45\,\% at typical room temperatures (around 25\,$^{\circ}$C) and with freshly prepared silane solutions.

\subsection{Coating Procedure}
Despite the negative implications of water for the silane solution, the presence of water is also necessary for a certain step of the SAM preparation: Silberzahn et al.~\cite{Silberzahn1991}, Angst and Simmons~\cite{Angst1991}, Tripp et al.~\cite{Tripp1992} and Wang et al.~\cite{Wang2003} highlighted that a thin layer (3\,nm) of water~\cite{Asay2005} adsorbed on the \ce{Si} wafer surface improves the quality of the resulting SAM drastically. Implementing this aspect in step (iv), the cleaned \ce{Si} wafer is taken out of the hot DI-water bath and then dried in a \ce{N2} jet. When no water is visible anymore on the surface, it is exposed shortly (half a second ---in order to prevent the formation of droplets) to the vapor of a boiling DI-water bath until a breath figure appears, which should evaporate again within a second. Immediately after the breath figure vanishes, the wafer is put for 15\,min into the silane solution, which should completely cover the polished side of the wafer. Afterwards, the wafer is taken out of the silane solution and flushed with a jet of chloroform out of a wash bottle to remove unbound silane molecules.

Especially on humid days, a white 'haze' can build up on the silane solution for the same reason mentioned before: silane molecules cross-link in the presence of water~\cite{Brzoska1994}. If this is the case, the surface of the solution should be skimmed with an extra piece of a cleaned \ce{Si} wafer. A similar haze can be observed sometimes on the silanized \ce{Si} wafer. It can be easily removed by thoroughly rinsing the wafer with chloroform.\\

Before placing the wafer back into the silane solution, all chloroform residues have to be removed by drying the entire wafer (both sides) with a jet of clean, dry \ce{N2}. Repeating the process of rinsing with chloroform and bathing in silane solution enhances the quality of the SAM. Typically, three iterations are sufficient.
With some experience, the quality of the SAM can already be judged by the way the chloroform and the silane solution bead off the surface: If there is a smooth, straight three phase contact line between liquid, solid and air, with a high contact angle and a low hysteresis (which can be judged by the trained eye), the silanization was successful and a homogeneous SAM can be expected. \\ 

In the following section, the resulting SAM surfaces are characterized in terms of their wetting properties, layer thickness and the surface roughness on different lateral length scales. A structural analysis using X-ray reflectometry of the SAM resulting from this procedure is presented in Ref.~\cite{Gerth}.

\section{Characterization}

\subsubsection{Water contact angles and surface energy} Following the preparation procedure presented in this study, advancing water contact angles of 114(2)$^{\circ}$ and a contact angle hysteresis below 10$^{\circ}$ can reliably be achieved for DTS, HTS and OTS (\textit{c.\,f.\ }Table~\ref{tableMessergebnisse}). The contact angles were obtained with fresh DI-water using the sessile drop method (OCA 20, Dataphysics Instruments GmbH, Filderstadt, Germany). The surface energy of the silane SAM can be approximated using the Good Girifalco relation~\cite{Good1960}, if the contact angles of two apolar liquids are known. Using 1-bromonaphtalene (> 95\,\%, Fluka) and bicylcohexyl (99\,\%, Acros Organics), contact angles on OTS of 40(4)$^{\circ}$ and 60(4)$^{\circ}$, respectively, were achieved. The results for the surface energy of the different silane SAMs are listed in Table~\ref{tableMessergebnisse}.

\subsubsection{Structural analysis} For the structural analysis of the SAMs, X-Ray reflectometry (XRR)~\cite{Gerth} was applied. By XRR, a layer thickness of 2.17(1)\,nm~\cite{Gerth} was measured for the OTS tail-group, which is in excellent agreement with the theoretically calculated length of the OTS tail in \textit{all-trans} configuration of 2.18\,nm~\cite{Gerth}, indicating an upright orientation. The results for DTS and HTS are compiled in Table~\ref{tableMessergebnisse} and are detailed in Ref.~\cite{Gerth}.

\subsubsection{Surface topography and atomic force microscopy (AFM)} The quality of the SAM in terms of surface coverage and roughness is characterized by atomic force microscopy (AFM, a Multimode and a FastScan Icon, both Bruker Nano Inc., Santa Barbara, CA, USA) in Tappingmode\texttrademark. 

\begin{figure}
  \includegraphics[width=0.48\textwidth]{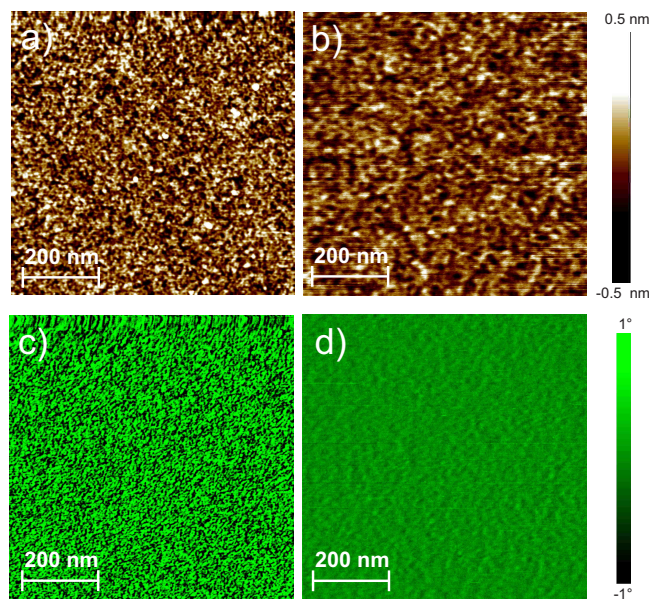}
  \caption{AFM images: Topography of an OTS-SAM (a) in comparison with a bare \ce{Si} wafer (b) prior to the preparation of the SAM and corresponding phase signal (c) of an OTS-SAM and (d) \ce{Si} wafer.}
  \label{imageOTS}
\end{figure}

\begin{figure}
  \includegraphics[width=0.47\textwidth]{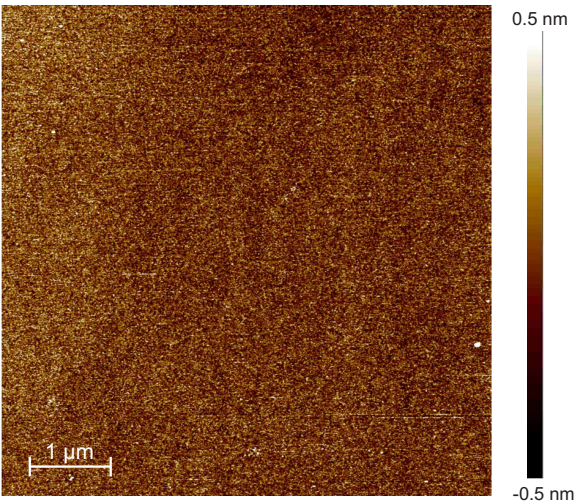}
  \caption{AFM height data of a representative (6\,$\upmu$m)$^2$ scan area (616 x 616 pixels) obtained from OTS. The image visualizes the large-scale homogeneity and full surface coverage of the SAM.}
  \label{imageOTS_gross}
\end{figure}

On a 1\,$\upmu$m$^2$ area, the root mean square (rms) roughness of the OTS layer is determined to 0.17(3)\,nm, that of a bare Si wafer to 0.14(2)\,nm (c.f. Figure~\ref{imageOTS} a and b, 515 x 512 pixel). Figure~\ref{imageOTS_gross} exemplifies a (6\,$\upmu$m)$^2$ AFM scan (616 x 616 pixel) of a typical OTS surface: The surface is homogeneously covered by the SAM and - at variance with other publications~\cite{Flater2007,Carraro1998}- no holes or grain boundaries can be detected. Operating the AFM in Tappingmode\texttrademark enables us to record an additional information, the phase lag between the free end of the cantilever and the driving piezo element ,which can be taken as a measure for the energy dissipation~\cite{Tamayo1998}. On a flat surface (rms < 1\,nm), differences of the phase lag typically hint to a chemically heterogeneous surface~\cite{Knoll2001}. Figure~\ref{imageOTS} c and d depict the phase signal for the OTS-SAM and a bare \ce{Si} wafer. On both surfaces, a very low phase lag (below 2$^{\circ}$) is recorded, which is an additional indication for the chemical homogeneity of the surface.

\begin{table*}[h]
\small
  \caption{Surface properties of silane SAMs.}
  \label{tableMessergebnisse}
  \begin{tabular}{c|cccccc}
    \hline
    silane&$\ce{H2O} \theta_{adv}$ & $\Delta\theta$ & surface&  thickness silane tail& tilt angle &roughness (rms)\\
    &&& energy& (XRR)~\cite{Gerth}&of the silane &1$\upmu m^2$ AFM scan\\
    & &  & / mNm$^{-1}$& / nm&tail~\cite{Gerth} &/ nm \\
    \hline
   DTS & 114(2)$^{\circ}$ & 5$^{\circ}$  & 26(1)  &1.31(3)& 22(1)$^{\circ}$&0.13(3)\\
   HTS & 114(2)$^{\circ}$ & 8$^{\circ}$  & 25(1)  &1.87(3)& 14(1)$^{\circ}$&0.16(2)\\
   OTS & 111(2)$^{\circ}$ & 5$^{\circ}$  & 24(1)  &2.17(1)&  5(1)$^{\circ}$&0.17(2)\\
  
    \hline
  \end{tabular}
\end{table*}

A similar high quality surface can be found for HTS and DTS (c.f. Table~\ref{tableMessergebnisse} and Ref.~\cite{Gerth}). Though the preparation temperature of 25\,$^{\circ}$C  is above the ideal coating temperature reported by Brzoska et al.~\cite{Brzoska1994} for the three types of silanes studied, the rms roughness of the different SAMs are similar (\textit{c.\,f.}\ Table~\ref{tableMessergebnisse}). The tilt angles of the hydrocarbon chains, however, are different as is detailed in the XRR study by Gerth et al.~\cite{Gerth} (\textit{c.\,f.}\ Table~\ref{tableMessergebnisse}). 

The following section compares the recipe presented here with the most common procedures reported in literature concerning time efficiency of the preparation procedure.
\section{Comparison}
Since the pioneering work of Zisman in 1946~\cite{Bigelow1946,Bigelow1947}, a variety of recipes to prepare monolayers of silanes via self-assembly on \ce{SiO2} substrates were reported.

\begin{table*}[h]
\small
  \caption{Comparison of the most common silanization recipes.}
  \label{tableVergleich}
  \begin{tabular}{l|llll}
    \hline
     & container & substrate  & silanisation  & finishing\\
     &cleaning &cleaning &solution&  treatment \\

    \hline

   Sagiv\cite{Sagiv1980}& not reported & 30\,min  & no time reported  & not reported \\
 (1980)& & (HNO$_3$, NaOH), &  (HCl$_3$, CCl$_4$, alkane) & \\

      \hline
   Wasserman\cite{Wasserman1989} & not reported & $\approx$ 1\,h  & 1-48\,h  & CH$_2$Cl$_2$, CHCl$_3$, \\
    (1989)& & (piranha solution) &  (hexadecane or  & ethanol \\
     & &  & bycyclohexyl, CCl$_4$) &  \\
 
 \hline	 
   Brzoska\cite{Brzoska1994} & not reported & several hours  & between 10\,min &  not reported\\
      (1994)&  & (piranha solution,  & and 1\,h(CCl$_4$, alkane) & \\
      &&\ce{O3} plasma)&&\\
      
       \hline
   Wang\cite{Wang2003} & 48\,h & $\approx$ 4\,h  & 18\,h to 19\,d  &  CH$_2$Cl$_2$,\\
     (2002)&  &(CHCl$_3$\, acetone,  & (isoprop G, acetone,& acetone, alcohol, \\
     &  &ethanol, RCA1, RCA2) &  CHCl$_3$)  & 12\,h in a \\
     &  &                     &               &desiccator\\

    \hline
   Mezger\cite{Mezger2008} &not reported & $\approx$ 1\,h & 3\,h  & hexadecane, \\
     (2008)& & (UV in O$_2$ & (hexadecane, CHCl$_3$) & toluene, 1\,h \\
      & & atmosphere and 10\,min&& at 100\,$^\circ\ $C\\
      && in piranha solution)&&\\

     \hline
   this study & $\approx$30\,min & around 1.5\,h& 3x15\,min (CCl$_4$,&rinsed with CHCl$_3$\\
       & (piranha solution) &  (piranha solution) &  bicyclohexyl), &\\
       &  &   & with CHCl$_3$ in between   &\\

    \hline
  \end{tabular}
\end{table*}

However, chemical cleaning of the \ce{Si} wafers always represents the first step of the SAM preparation. Based on either alkaline or acidic solutions, various procedures are reported in literature: Wang et al.~\cite{Wang2003} applied RCA1 and RCA2~\cite{Kern1970}, two procedures well established in the semiconductor industry. Sagiv~\cite{Sagiv1980} used nitric acid, Wasserman et al.~\cite{Wasserman1989} and Mezger et al.~\cite{Mezger2008} as well as this recipe used piranha solution to clean the \ce{SiO2} surfaces. A ozone plasma treatment is reported by Brzoska et al.~\cite{Brzoska1994}. Most recipes involve rinsing of the substrates with DI-water to remove aggressive cleaning chemicals.

An important feature to our experience is the cleaning of the containers and tweezers, which, however, is described only by a few articles, \textit{e.\,g.}\ Wang et al.~\cite{Wang2003} and this study.

In terms of the chemical composition of the silane solution, different options can be found in the literature. The most common solvents for the silane solution are: hexadecane~\cite{Wasserman1989,Brzoska1994}, Isopar G~\cite{Wang2003} or - as in this recipe - bicyclohexyl~\cite{Wasserman1989,Bierbaum1995}. To avoid the agglomeration of micelles caused by water in the solution, small amounts of either chloroform or carbon tetrachloride are added. 

The time period the \ce{Si} wafer is exposed to the silane solution differs significantly from recipe to recipe: typical exposure times range from several minutes~\cite{Brzoska1994} to 10 days~\cite{Wang2003}. Afterwards, the SAMs are rinsed with a solvent (chloroform~(\ce{CHCl3})~\cite{Wasserman1989} or dichloromethane (\ce{CH2Cl2})~\cite{Wasserman1989,Bierbaum1995}) to remove unbound silanes. As described above, the silane coated substrates are rinsed with chloroform several times during the coating process, which improves the quality of the resulting SAM. 

Several articles describe different kinds of post-treatment of the prepared SAMs to enhance the cross-linking of the silanes such as heating in an oven~\cite{Bierbaum1995} or storage in a desiccator~\cite{Wang2003}. The recipe presented here features only a rinse in chloroform as the last step. The coated samples can then be stored in a dust-controlled environment in air. As compared to \ce{Si} wafers, the silanized wafers exhibit a very low surface energy. Physisorbed particles, generated for instance by cutting the wafers, can usually be flushed away easily, \textit{e.\,g.}\ by sonicating the samples in polar and apolar ultra-clean solvents like ethanol, acetone or toluene~\bibnote{Especially for toluene, we recommend using semiconductor industry-grade quality for its particle-controlled content, for instance LiChroSolv\textsuperscript{\textregistered} (Sigma-Aldrich) or Optima\textsuperscript{\textregistered} (Fisher Scientific).} for a few minutes.

\section{Conclusion}
In conclusion, we present a reliable and easy-to-follow recipe that leads to silane SAMs of high quality in terms of a low contact angle hysteresis and a very low surface roughness (comparable to a bare \ce{Si} wafer), both resulting from the excellent topographical and chemical homogeneity of the SAMs. The characterization of the SAMs includes AFM, contact angle measurements, ellipsometry and X-ray reflectometry~\cite{Gerth}. In addition, it is optimized concerning preparation time and allows the silanization of a complete \ce{Si} wafer in about 3\,h. The throughput can be enhanced, if the cleaning and coating process is simultaneously performed for several wafers. With some practice and sufficient amount of clean beakers, 3-4 wafers can be prepared in one run, which corresponds to an average preparation time of 45 to 60\,min per wafer. The recipe has been tested in several labs with different environmental conditions with similar experience and similar quality. With this study, we like to set a lab standard for commonly used silane coatings on \ce{Si} wafers at normal lab conditions. The intention is to facilitate a comparison of silane coatings of different studies and different labs. Standard quality checks of a silane SAM should comprise at least an advancing and receding water contact angle measurement as well as the determination of the sample topography of a 1-10\,$\upmu$m-sized area and roughness before and after coating.  

\bibliography{Silanrezept}

\begin{mcitethebibliography}{58}
\bibitem[Not()]{Note-1}
It is essential to use non-stabilized \ce{H2O2}.\relax
\mciteBstWouldAddEndPunctfalse
\mciteSetBstMidEndSepPunct{\mcitedefaultmidpunct}
{}{\mcitedefaultseppunct}\relax
\EndOfBibitem
\bibitem[Not()]{Note-2}
For handling the chemicals, local safety rules apply and the use of lab coats,
  goggles and suitable gloves is mandatory.\relax
\mciteBstWouldAddEndPunctfalse
\mciteSetBstMidEndSepPunct{\mcitedefaultmidpunct}
{}{\mcitedefaultseppunct}\relax
\EndOfBibitem
\bibitem[Not()]{Note-4}
For the silanization of a 4\,inch wafer, three PTFE beakers of similar
  diameter~\bibnote{It is convenient to choose the diameter of the beaker only
  slightly larger than the diameter of the wafer as this allows a complete
  coverage of the wafer with a minimum amount of piranha solution and silane
  solution required.}, one 2\,l Erlenmeyer flask for the hot DI-water, one
  graduated cylinder (50\,ml) and pipettes for the chemicals of the
  silanization solution.\relax
\mciteBstWouldAddEndPunctfalse
\mciteSetBstMidEndSepPunct{\mcitedefaultmidpunct}
{}{\mcitedefaultseppunct}\relax
\EndOfBibitem
\bibitem[Not()]{Note-5}
Crystal orientation and type of doping of the \ce{Si}-wafers do not impact on
  the silanization\relax
\mciteBstWouldAddEndPuncttrue
\mciteSetBstMidEndSepPunct{\mcitedefaultmidpunct}
{\mcitedefaultendpunct}{\mcitedefaultseppunct}\relax
\EndOfBibitem
\bibitem[Not()]{Note-7}
Especially for toluene, we recommend using semiconductor industry-grade quality
  for its particle-controlled content, for instance
  LiChroSolv\textsuperscript{\textregistered} (Sigma-Aldrich) or
  Optima\textsuperscript{\textregistered} (Fisher Scientific).\relax
\mciteBstWouldAddEndPunctfalse
\mciteSetBstMidEndSepPunct{\mcitedefaultmidpunct}
{}{\mcitedefaultseppunct}\relax
\EndOfBibitem
\bibitem[Not()]{Note-8}
It is convenient to choose the diameter of the beaker only slightly larger than
  the diameter of the wafer as this allows a complete coverage of the wafer
  with a minimum amount of piranha solution and silane solution required.\relax
\mciteBstWouldAddEndPunctfalse
\mciteSetBstMidEndSepPunct{\mcitedefaultmidpunct}
{}{\mcitedefaultseppunct}\relax
\EndOfBibitem
\end{mcitethebibliography}

\acknowledgement
We gratefully acknowledge financial support from German Science Foundation DFG under Grant No. Ja 905/3 within the priority program 1164, Grant No. Se 1118/2 and the graduate school GRK 1276 for financial support.

\end{document}